\documentclass[aps,twocolumn,showpacs,english]{revtex4}

\usepackage[T1]{fontenc}
\usepackage[latin9]{inputenc}
\usepackage{textcomp}
\usepackage{amsmath}
\usepackage{graphicx}
\usepackage{amssymb}
\usepackage{babel}

\begin{document}

\title{Extraction of the homogeneous linewidth of a fast spectrally diffusing line}

\author{S. Bounouar$^{1,2}$, A. Trichet$^{1}$, M. Elouneg-Jamroz$^{1,2}$, R. Andr\'{e}$^{1}$, E. Bellet-Amalric$^{2}$, C. Bougerol$^{1}$, M. Den Hertog$^{3}$, K. Kheng$^{2}$, S. Tatarenko$^{1}$,  and J.-Ph.~Poizat$^{1}$}

\affiliation{
$^1$ CEA-CNRS-UJF group 'Nanophysique et Semiconducteurs', Institut N\'{e}el, CNRS - Universit\'{e} Joseph Fourier, 38042 Grenoble, France, \\
$^2$ CEA-CNRS-UJF group 'Nanophysique et Semiconducteurs', CEA/INAC/SP2M, 38054 Grenoble, France, \\
$^3$ Institut N\'{e}el, CNRS - Universit\'{e} Joseph Fourier, 38042 Grenoble, France,}

\begin{abstract}
We present a simple method to extract the homogeneous linewidth of a single photon emitter line exhibiting fast (down to 1 ns) spectral diffusion (SD). It is based on a recently developed technique using photon correlation measurements on half of the line. Here we show that the SD induced bunching depends on the ratio between the width of the homogeneous line and the spectral diffusion amplitude. Using this technique on a CdSe/ZnSe quantum dot,
we investigate the temperature dependence of its fast SD amplitude and its homogeneous excitonic linewidth.



\end{abstract}


\maketitle

Spectral diffusion (SD) corresponds to random spectral jumps of a narrow line within a broader spectral profile.
It is a prominent issue in spectroscopy \cite{Klauder,Plakhotnik,Empedocles,Ambrose,Robinson,Seufert,Turck,Besombes}. From a practical point of view, it broadens the observed linewidth and prevents the access to the intrinsic line properties. On a more fundamental side, it gives some very local information on the microscopic environment of a single light emitter embedded in a solid matrix, or moving within a fluid or a gas.
The quest for large light matter coupling in condensed matter physics makes use of photonic nanostructures such as photonic crystals \cite{Englund,vuckovic}, photonic wires \cite{Claudon,Yeo} or metallic nanoantenae \cite{Novotny} where the emitter is not far from a surface and may therefore undergo spectral diffusion caused by surface charge fluctuations.

 Visualizing directly the spectral wandering
  by recording a time series of spectra
  has been so far the usual method to observe SD \cite{Plakhotnik,Empedocles,Ambrose,Robinson,Seufert,Turck,Besombes}.
 For single photon emitters, the time resolution was therefore limited by the minimum time of about 1 ms required for a
photon counting charged coupled device (CCD) to acquire a spectrum.

Except from the exotic possibility of  statistics such as Levy flight \cite{Brokmann}, SD is usually entirely described by its fluctuation amplitude and its characteristic time. But, additionally, one might wish to access  the width of the homogeneous wandering line. When the characteristic SD time is faster than the time necessary to record a spectrum, this information is hidden by the inhomogeneous linewidth and difficult to access for a single photon emitter.
Sophisticated methods featuring both high spectral and temporal resolution are required to access this quantity.
To our knowledge, photon correlation Fourier spectroscopy (PCFS) \cite{Coolen} is the only published method able to extract the homogeneous linewidth of a rapidly wandering line of a single emitter together with its SD parameters. In PCFS, high spectral resolution ($\Delta \lambda /\lambda = 10^{-6}$) is provided by a Michelson interferometer, and  temporal resolution of  20 $\mu$s is achieved with a photon correlation set-up  at the outputs of the interferometer.
Alternatively, four wave mixing \cite{Langbein} is able to give  the coherence time of the transition of a single emitter undergoing spectral diffusion, but without any access to SD characteristic time.
In the case of an ensemble of emitters, lock-in hole burning techniques \cite{Palinginis} can give the homogeneous linewidth.

\begin{figure} 
\resizebox{0.75\textwidth}{!}{\includegraphics{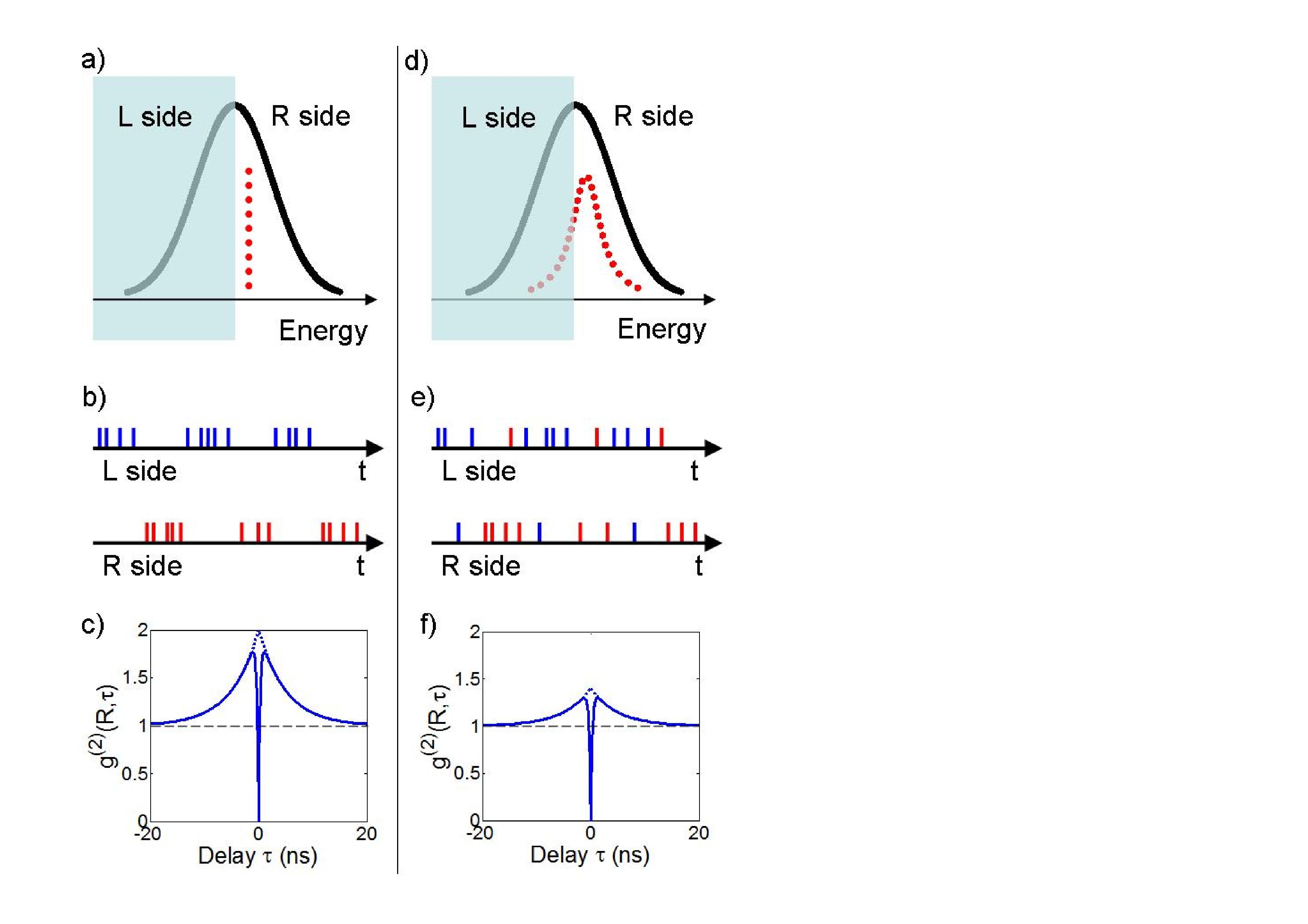}}
\caption{(Color online) Principle of the PCS method in the case of infinitely narrow line (a-c) and in the case of a line of finite width (d-e). In a) and d) are represented the spectra of a spectrally diffusing single emitter with an infinitely narrow linewidth (in a)) and a finite linewidth (in d)). Only the right (R) spectral window is selected for detection. This corresponds to the photon statistics of the R side as shown in b) and e) and to the half line autocorrelation functions displayed as a solid line in c) and f). In b) and e), "red photons" correspond to photons emitted from a line whose center is located on the R side. However, in e), owing to the finite linewidth of the homogeneous line, red (resp. blue) photons can be emitted on the L (resp. R) side, leading to a reduction of the SD induced bunching. In c) and f), the zero delay dip is due to the single photon type antibunching and the dotted line is the bunching contribution caused by spectral diffusion only.
}\label{infiniteline}
\end{figure}

Our recently developed photon correlation spectroscopy (PCS) technique \cite{Sallen} converts spectral fluctuations into intensity fluctuations, as also reported in \cite{Zumbusch,Coolen,Marshall}. This simple method benefits  from the subnanosecond time resolution of an Hanbury-Brown and Twiss photon correlation set-up and the spectral resolution of the spectrometer.   In short, it is based on correlations of photons emitted within a spectral window narrower than the SD broadened line. Owing to the wandering of the homogeneous line, the emission energy stays a limited time within this spectral window leading to photon bunching (see fig \ref{infiniteline}). The characteristic time $\tau_c$ of this effect can be easily accessed by photon correlation even though it is much shorter than the inverse photon count rate.
In practice, spectral diffusion in the 1 ns range can be measured even though the count rate is less than one photon every 10 $\mu$s.
This opens a new domain of fast spectral diffusion phenomena that was previously not accessible.

In this letter, we present a development of the PCS technique that enables the extraction of the homogeneous linewidth of the spectrally diffusing line of a single photon emitter, together with its complete SD parameters. The method is performed on CdSe  quantum dots (QDs) embedded in a ZnSe nanowire (NW) \cite{JAP}, with a single object photoluminescence linewidth between  1 and 2 meV \cite{BounouarPRB}. After describing the principle of our measurement, we discuss the properties of the half-line autocorrelation function (HLAF) of a spectrally diffusing single photon emitter with a finite homogeneous linewidth. We show that the contrast of the measured photon bunching is related to the ratio between the  SD amplitude and the homogeneous linewidth of the wandering line. We evaluate these two quantities separately on our NW-QD and study their behavior when temperature is increased
as an illustration of the possibilities offered by our method.
This allows us to probe the emitter energy statistic as a function of temperature.

In the case of an infinitely narrow homogeneous line as shown in fig.\ref{infiniteline}(a-c), the HLAF, noted
$g^{(2)}(R,\tau)$ when the R side of the inhomogeneous line is detected, writes
\begin{equation}
g^{(2)}(R,\tau)=[1+\beta \exp(-\gamma_{c}\tau)][1-\exp(-(r+\gamma)\tau)],
\end{equation}
with $\gamma_{c}=\frac{1}{\tau_c}$ being the SD correlation rate, $\gamma$ the radiative emission rate, $r$ the pumping rate and $\beta$ the SD induced bunching factor.  This expression is the product of two terms.
The last term corresponds to the antibunching of the single photon emitter.
The first term accounts for  spectral diffusion  and its influence on the
ability of the emitter to send photons in the R spectral window.
Note that $\beta=1$ in the case of an infinitely narrow homogeneous line when the two energy sides are of equal importance (ie half of the total count in the R side). Equal distribution between L and R is checked very carefully since deviation from this  situation  leads to a different value of $\beta$ \cite{Sallen}.

We now consider the case of a finite homogeneous linewidth (fig.\ref{infiniteline}(d-f)). In this situation,
 even if the homogeneous line is centered in the L side, a photon can nevertheless be detected in the R side and vice versa as represented in fig. \ref{infiniteline}(e).
 This randomizes the photon stream   and results in a reduction of the bunching contrast $\beta$
  as seen in fig.\ref{infiniteline}(f).



A probabilistic model was developed \cite{theorsd} considering the consequence of a finite linewidth on the HLAF. The key idea is to take into account the combined effects of the correlated energy statistics of the SD together with the uncorrelated energy statistics introduced by the homogeneous linewidth. When the photons are detected on the R side, the bunching contrast $\beta$ is then found as
\begin{equation}
\beta=\frac{(\alpha_{in}-\alpha_{out})^{2}}{(\alpha_{in}+\alpha_{out})^{2}},
\label{beta}
\end{equation}
where $\alpha_{in,out}$ is a normalized overlap integral defined  as
\begin{equation}
\alpha_{in,out}=
\int_{\mu\epsilon R,L}\left[ D_\Sigma(\mu)\int_{E\epsilon R} H_\sigma(E-\mu)dE \right]d\mu ,
\label{eqalpha}
\end{equation}
where $D_\Sigma (\mu)$  is the SD line distribution of full width at half maximum (FWHM) $\Sigma$, i.e.  the probability density associated with the position of the homogeneous line median energy  $\mu$, and
$H_\sigma (E-\mu)$ the  homogeneous line shape with median $\mu$ and FWHM $\sigma$.
We assume here that each of these function has a given shape that can be described by a single width parameter ($\Sigma$ for $D_\Sigma$ and $\sigma$ for $H_\sigma$).
The coefficients $\alpha_{in}$ and $\alpha_{out}$ are the probability to detect a photon in the R spectral window with the homogeneous linewidth respectively centered in R or in L. They only depend on  the homogeneous and SD line profiles $H_\sigma$ and  $D_\Sigma$.
If the a priori knowledge of the physical system allows us to chose the shape of the SD distribution and of the homogeneous line, which is generally the case, the bunching contrast $\beta$ will only depend on their linewidth  ratio $\sigma/\Sigma$. This is the main result of the paper.

In the following, we have used a Lorentzian shape of FWHM $\sigma$ for   $H_\sigma$  and a Gaussian shape of FWHM $\Sigma$ for   $D_\Sigma$ to model the bunching factor $\beta$ and fit the experimental data. We are aware that a Lorentzian shape is only an approximation of the real homogeneous line shape, especially  at elevated temperature when the exciton-phonon coupling leads to more complex line shapes \cite{Besombes_line}, but this gives nevertheless a meaningful  estimation of the homogeneous linewidth and its variation.
 The Gaussian shape of the spectral diffusion distribution
 is justified by the Kubo-Anderson theory which describes the spectral diffusion as the result of fluctuations of a very large number of independent identical random variables, leading globally to a Gaussian distribution of the emitter energies in the "slow fluctuations" regime where $\Sigma \tau_c \gg \hbar$ \cite{kubo,anderson,Berthelot}. In our case $\Sigma\approx 1 $ meV and $\tau_c \approx 10$ ns so that $\Sigma \tau_c \approx 10^{4} \hbar$.

\begin{figure}
\resizebox{0.4\textwidth}{!}{\includegraphics{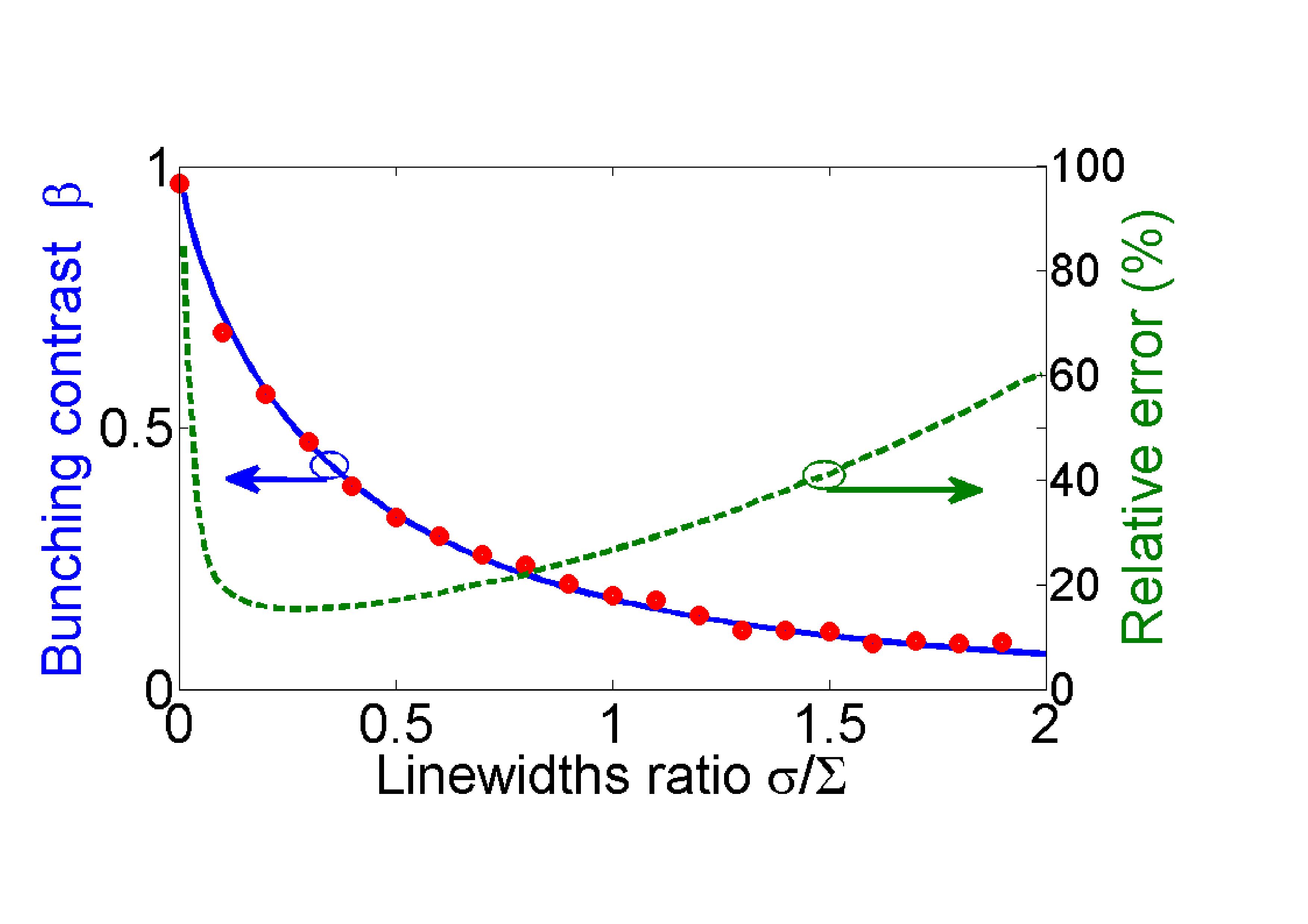}}
 \caption{(Color online) The blue solid line is the theoretical value of the bunching factor $\beta$ (cf eq (\ref{beta})) as a function of the linewidths ratio $\sigma/\Sigma$ in the case of a Gaussian inhomogeneous line with FWHM $\Sigma$ and a Lorentzian homogeneous line with FWHM $\sigma$. This result is confirmed  by a Monte-Carlo simulation (red dots).
 The green dashed line is the relative error $\delta (\sigma/\Sigma)/(\sigma/\Sigma)$ for the actual experimental error $\delta \beta = 0.04$.
 }
 \label{bunching}
\end{figure}

Within the above assumptions, the bunching factor $\beta$,
computed from eq.(\ref{beta}), is plotted in fig.\ref{bunching} as a function of $\sigma/\Sigma$ together with a Monte-Carlo simulation. The bunching factor decreases with this ratio because the uncorrelated statistics introduced by the homogeneous linewidth  is degrading the  bunching induced by the correlated SD process. Note that for an infinitely small homogeneous linewidth the bunching factor is equal to unity ($\alpha_{in}=1/2$ and $\alpha_{out}=0$), the energy of the emitter being only defined by SD. On the contrary, for an homogeneous linewidth larger than the SD amplitude the bunching vanishes
($\alpha_{in}\rightarrow1/4$ and $\alpha_{out} \rightarrow 1/4$), as the uncorrelated homogeneous linewidth process dominates and the emitter energy statistics in the time domain tends to be Poissonian.
As seen with the green dashed line in fig.\ref{bunching}, a relative error on $\sigma / \Sigma$ better than 30$\%$ is obtained for $0.04\leq \sigma / \Sigma \leq 1.1$ in our experimental conditions.

\begin{figure}
\resizebox{0.55\textwidth}{!}{\includegraphics{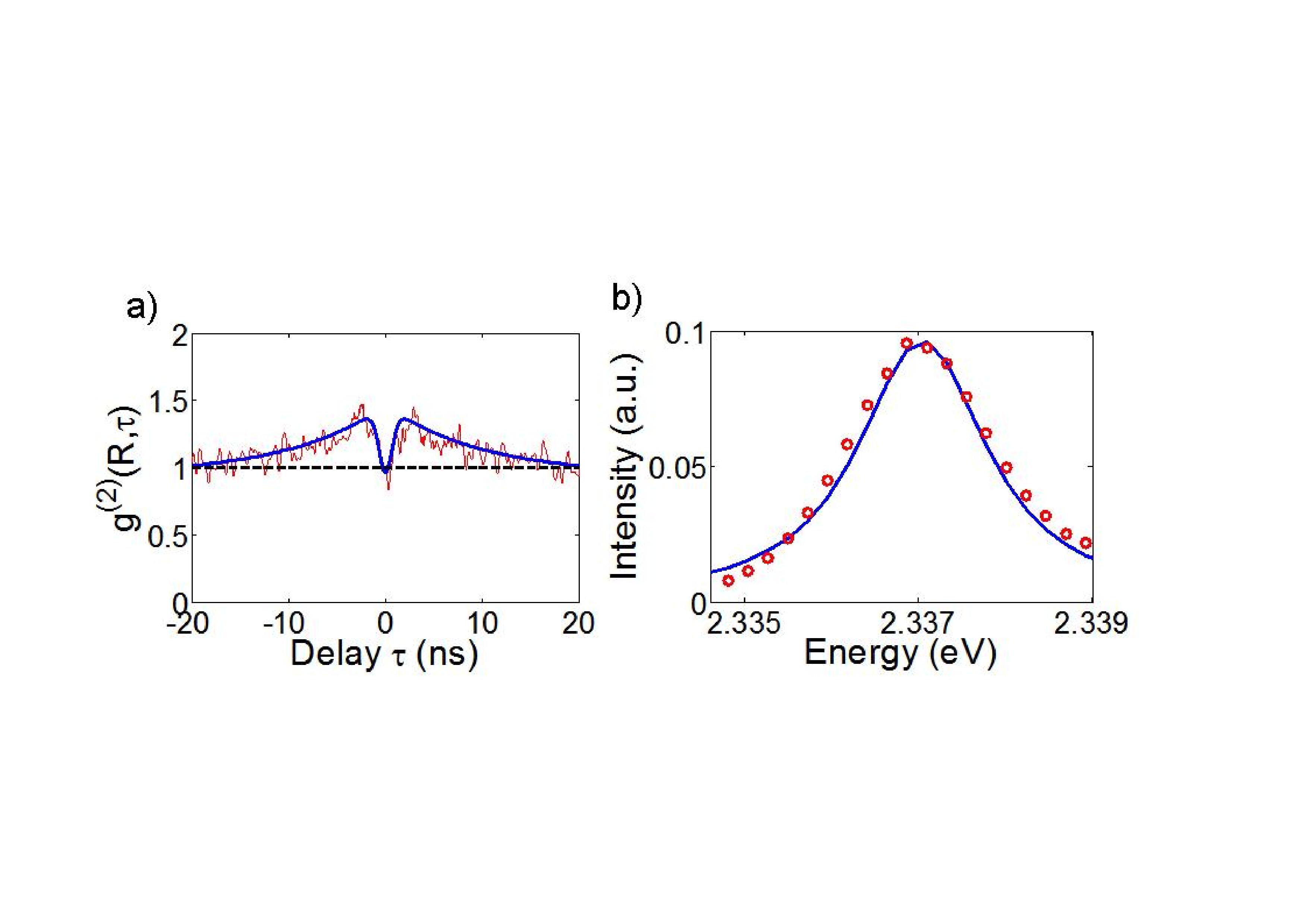}}
 \caption{(Color online) (a) Typical measured HLAF at 4K together with its best fit. The limited depth of the central antibunching dip is due to the limited time resolution of the photon correlation set-up.  (b) Emission line spectrum fitted with a Voigt profile.
  }
 \label{bunchspectrum}
\end{figure}

The measurement of the HLAF gives access to the bunching factor and therefore to the  $\sigma / \Sigma$ ratio.
The spectrum also depends on these two parameters as it is a convolution of the homogeneous linewidth  and of the spectral diffusion distribution. It leads to a Voigt profile when the homogeneous and the inhomogeneous line are respectively Lorentzian and Gaussian as  mentionned above.
Thus, the recording of the spectrum and of the HLAF (cf fig.\ref{bunchspectrum}) allows the extraction of the homogeneous linewidth $\sigma$ together with the SD amplitude $\Sigma$ and characteristic time $\tau_c$ in the 10 ns range.

Our experimental results have been obtained with  a time resolved microphotoluminescence
setup. The excitation source is a continuous wave laser emitting at $\lambda =405$ nm (ie $3.06$ eV).
The QD luminescence is detected through a $\delta \lambda=0.06$ nm (ie
$\delta E=300$ $\mu$eV) resolution spectrometer on a charged coupled device (CCD)
camera for spectra or on a Handbury-Brown and Twiss photon correlation set-up of overall time resolution of 800 ps (FWHM).

The measurement method was applied on three different QDs with different SD amplitudes and the extracted homogeneous linewidths were around 300 $\mu eV$ for the three measurements. This actually corresponds to the experimental spectral resolution imposed by the spectrometer. This is the main limitation of this method. However, we underline the fact that the measured correlation times are of the order of 10 ns and that it is the only reported method able to probe experimentally the homogeneous linewidth of such a fast diffusing single photon emitter.

\begin{figure}
\resizebox{0.5\textwidth}{!}{\includegraphics{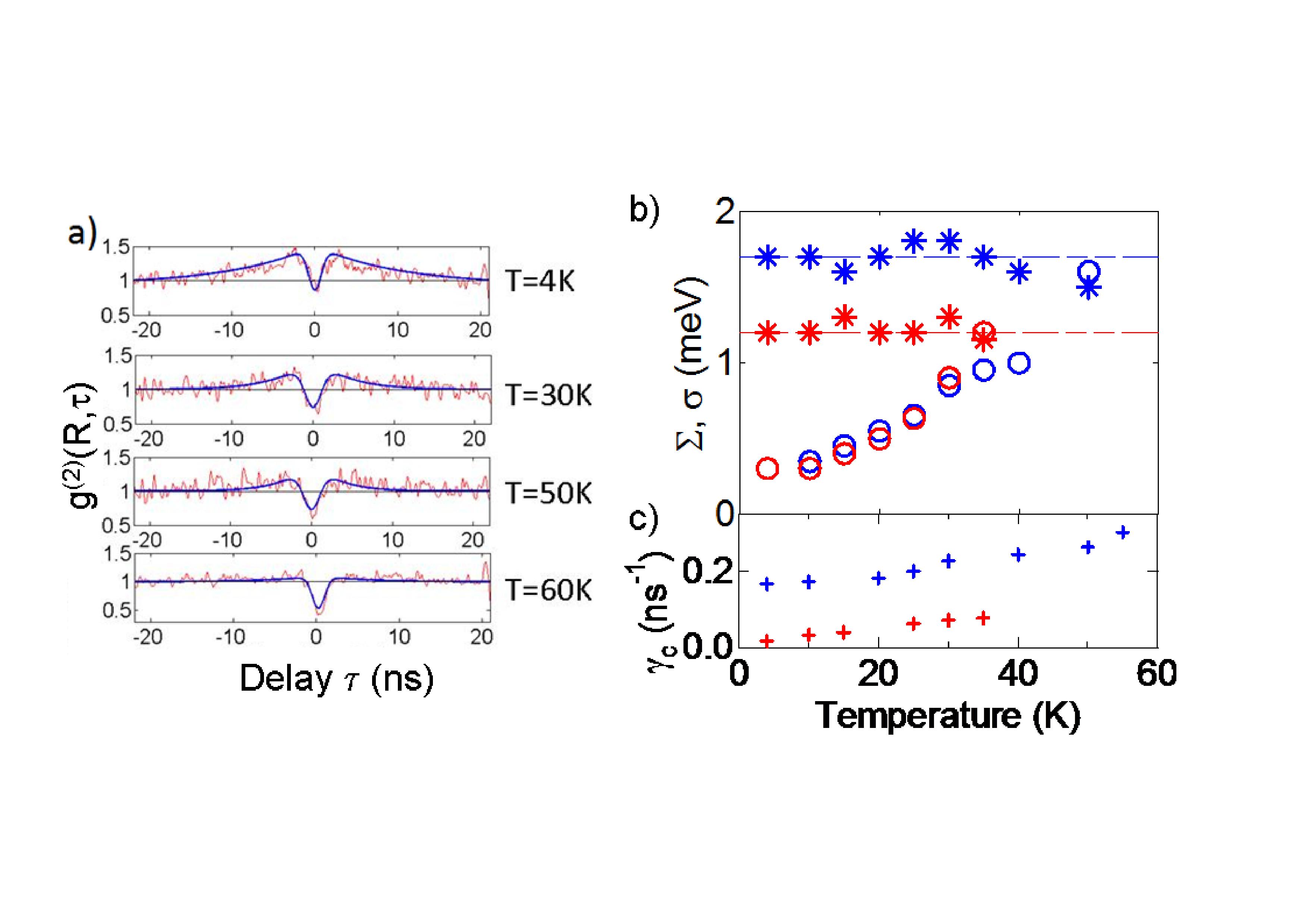}}
 \caption{(Color online) a) HLAF for different temperatures for QD1 with the corresponding fit. b) Temperature dependency of the homogeneous linewidth ($\circ$) and SD amplitude ($\ast$) extracted from HLAF and linewidth measurements for QD1 (in blue) and QD2 (in red).
 c) Temperature dependency of the SD rate $\gamma_c$ for QD1 (in blue) and QD2 (in red).}
 \label{fig:sample}
\end{figure}

When the temperature is increased, the first noticeable consequence on the HLAF is the decrease of the bunching contrast (fig.\ref{fig:sample} a)). The first interpretation of this observation is the increase of the homogeneous linewidth owing to temperature induced phonon broadening \cite{Besombes_line,Favero}. This process becomes dominant above 60 K for this QD, and the bunching  collapses completely for higher temperatures meaning that the  $\sigma/\Sigma$ ratio becomes large.
 At high temperature the energy of the emitter is less and less time correlated and tends to adopt a Poissonian statistics.
 We display the extracted homogeneous linewidths  and spectral diffusion amplitudes (fig.\ref{fig:sample} b))
 together with the spectral diffusion rate $\gamma_c=1/\tau_c$ (fig.\ref{fig:sample} c))
 as a function of temperature for two different QDs  with different $\Sigma$ (blue: QD1,  $\Sigma=1.7$ meV and red: QD2,  $\Sigma=1.2 $meV). The effective homogeneous linewidths for both QDs increase with temperature. They reach the same order of magnitude as the fluctuation amplitude  for temperatures around 40-50 K. For temperatures
higher than  60 K, the bunching becomes no longer measurable and the $\sigma / \Sigma$ ratio can no longer be accurately  extracted.

A remarkable  feature that appears  in fig.\ref{fig:sample} b) is that the SD amplitude
is constant with temperature for the two investigated quantum dots.
So our results show that the main contribution to the QD linewidth is  SD for temperatures below 60 K and  phonon broadening above.
This relatively surprising behaviour can be explained with the help of the Kubo Anderson model adapted to non symmetrical fluctuations \cite{Cook, Berthelot}. Even though the description of spectral diffusion mechanisms in QDs is not the main focus of this paper, we give here a brief interpretation of this result.  Emission energy fluctuations in  semiconductors are caused by captures and escapes of carriers in defects in the QD vicinity.
 These carrier motions are mainly governed by phonon  and Auger effects.
 The Auger  contribution is similar for  capture and  escape processes and
is dominant at  high pumping power.
This means that, under our experimental conditions, $\gamma_{\uparrow} \approx \gamma_{\downarrow}$
whatever the temperature.
Since the SD amplitude is governed by $\gamma_{\uparrow} / \gamma_{\downarrow}  $ \cite{Berthelot}, it
 should therefore be temperature independent, as observed experimentally. Furthermore this effect is rather robust to a departure of the $\gamma_{\uparrow} / \gamma_{\downarrow}  $ ratio from unity owing to the fact that the SD amplitude features a rather flat maximum around $\gamma_{\uparrow} / \gamma_{\downarrow}  =1$.
Fig.\ref{fig:sample} c) shows that the SD rate $\gamma_c=1/\tau_c$ increases with temperature. This rate is given by
$\gamma_c=\gamma_{\downarrow}+\gamma_{\uparrow}$ \cite{Berthelot}.
Its rise  with temperature is caused by the rise of $\gamma_{\downarrow}$ and $\gamma_{\uparrow}$ with the increasing phonon population.
The difference between the two QDs  can be attributed to a difference in the Auger rate and/or to a different defect energy depth.

As a conclusion, we have demonstrated that  the measurement of the HLAF brings informations on the energy statistics of the emitter in addition of providing a subnanosecond resolution access to the correlation time of the fluctuations.
The main result of this paper is the possibility
of  extracting of the homogeneous linewidth of a  fast diffusing line with a simple photon correlation set-up.
This work brings a new tool for
 the understanding of single emitters environment fluctuations.
As an illustration, we showed that, in CdSe/ZnSe NW QDs, SD amplitude does not depend on temperature and that the line broadening is only due to the coupling of the emitter with phonons.  This
 contributes to a better   interpretation of the temperature behaviour  of QD linewidths.

We acknowledge support from the French National Research Agency (ANR) through the Nanoscience and Nanotechnology Program (Project BONAFO ANR-08-NANO-031-01) that provided a research fellowship for MdH. MEJ acknowledges financial support from the Nanosciences Foundation "Nanosciences, aux limites de la nano\'{e}lectronique" (RTRA).

\begin{thebibliography}{10}


\bibitem{Klauder} J. R. Klauder and P. W. Anderson,
Phys. Rev. \textbf{125}, 912 (1962)



\bibitem{Ambrose}  W. P. Ambrose, and W. E. Moerner,
Nature \textbf{349}, 225 (1991).

\bibitem{Empedocles} S. A. Empedocles, D. J.  Norris, and M. G. Bawendi,
Phys. Rev. Lett. \textbf{77}, 3873 (1996).

\bibitem{Robinson} H. D. Robinson and B. B. Goldberg,
Phys. Rev. B  \textbf{61},  R5086 (2000).

\bibitem{Plakhotnik} T. Plakhotnik and D. Walser, Phys. Rev. Lett. \textbf{80}, 4064 (1998)

\bibitem{Seufert}
J. Seufert, R. Weigand, G. Bacher, T. K\"{u}mmell, A. Forchel, K. Leonardi, and D. Hommel
Appl. Phys. Lett. \textbf{76}, 1872 (2000)

\bibitem{Turck}
V. T\"{u}rck, S. Rodt, O. Stier, R. Heitz, R. Engelhardt, U. W. Pohl, D. Bimberg, and R. Steingruber,
Phys. Rev. B  \textbf{61}, 9944 (2000).

 \bibitem{Besombes}
L. Besombes, K. Kheng, L. Marsal, and H. Mariette,
Phys. Rev. B \textbf{65}, 121314 (2002).


\bibitem{Englund}
D. Englund, A. Faraon, I. Fushman, N. Stoltz, P. Petroff and J. Vu\v{c}kovi\'{c},
Nature (London) \textbf{450}, 857 (2007)


\bibitem{vuckovic}
A. Majumdar, E. D. Kim, and J. Vu\v{c}kovi\'{c}, Phys. Rev. B \textbf{84}, 195304 (2011)

\bibitem{Claudon}
J. Claudon, J. Bleuse, N.S. Malik,   M. Bazin,    P. Jaffrennou,    N. Gregersen,
    C. Sauvan,  P. Lalanne and  J.-M. G\'{e}rard,
  Nature Photon. \textbf{4}, 174 (2010).

\bibitem{Yeo}  I. Yeo, N.S. Malik, M. Munsch, E. Dupuy, J. Bleuse, Y.-M. Niquet, J.-M. G\'{e}rard, J. Claudon,
E. Wagner, S. Seidelin, A. Auff\`{e}ves, J.-Ph. Poizat, and G. Nogues,
    Appl. Phys. Lett. \textbf{99}, 233106 (2011)

\bibitem{Novotny}
L. Novotny and N. van Hulst, Nature Photon. \textbf{5}, 83 (2011)


\bibitem{Brokmann}
X. Brokmann, J.-P. Hermier, G. Messin, P. Desbiolles, J.-P. Bouchaud, and M. Dahan, Phys. Rev. Lett. \textbf{90}, 120601 (2003)

\bibitem{Coolen}
X. Brokmann, M. Bawendi, L. Coolen, J.-P. Hermier,
Opt. Express, \textbf{14}, 6333 (2006);
L. Coolen, X. Brokmann, and J.-P. Hermier, Phys. Rev. A \textbf{76}, 033824 (2007);
L. Coolen, X. Brokmann, P. Spinicelli, and J.-P. Hermier
Phys. Rev. Lett. \textbf{100}, 027403 (2008).

\bibitem{Langbein}
B. Patton, W. Langbein, U. Woggon, L. Maingault and H. Mariette,
Phys. Rev. B \textbf{73}, 235354 (2006);
W. Langbein and B. Patton,
J. Phys. Condens. Matter \textbf{19} 295203 (2007);
W. Langbein, Rivista  Nuovo Cimento  \textbf{33}, 255 (2010)


\bibitem{Palinginis} P. Palinginis, S. Tavenner, M. Lonergan, and H. Wang
Phys. Rev. B \textbf{67}, 201307 (2003).

\bibitem{Sallen} G. Sallen, A. Tribu, T. Aichele, R. Andr\'{e}, L. Besombes, C. Bougerol, M. Richard, S. Tatarenko, K. Kheng, and J. Ph.~Poizat, Nature Photon. \textbf{4}, 696 (2010)
 	
\bibitem{Zumbusch} A. Zumbusch, L. Fleury, R.  Brown, J.  Bernard, and M.  Orrit,
Phys. Rev. Lett. \textbf{70}, 3584 (1993).


\bibitem{Marshall}
L. F. Marshall, J. Cui, X. Brokmann, and M. G. Bawendi,
Phys. Rev. Lett. \textbf{105}, 053005 (2010).



\bibitem{JAP} M. Den Hertog, M. Elouneg-Jamroz, E. Bellet-Amalric, S. Bounouar,
    C. Bougerol, R. Andr\'{e}, Y. Genuist, J.-Ph. Poizat, K. Kheng and S. Tatarenko,
 J. Appl. Phys. \textbf{110}, 034318 (2011).

 \bibitem{BounouarPRB} S. Bounouar, C. Morchutt, M. Elouneg-Jamroz, L. Besombes, R. Andr\'{e}, E. Bellet-Amalric, C. Bougerol, M. Den Hertog, K. Kheng, S. Tatarenko,  and J. Ph.~Poizat,
     Phys. Rev. B \textbf{85}, 035428 (2012).

\bibitem{theorsd}
A. Trichet and S. Bounouar.
arXiv:1201.4016

\bibitem{Besombes_line} L. Besombes, K. Kheng, L. Marsal, and H. Mariette
Phys. Rev. B \textbf{63}, 155307 (2001).


\bibitem{kubo}R. Kubo,
J. Phys. Soc. Jap. \textbf{9}, 935 (1954).

\bibitem{anderson}P. W. Anderson,
J. Phys. Soc. Jpn. \textbf{9}, 316 (1954).

\bibitem{Berthelot} A. Berthelot, I. Favero, G. Cassabois, C. Voisin, C. Delalande, Ph. Roussignol, R. Ferreira,
and J.-M. G\'{e}rard,
Nature Physics \textbf{2},  759  (2006).

\bibitem{Favero}
I. Favero, A. Berthelot, G. Cassabois, C. Voisin, C. Delalande, Ph. Roussignol, R. Ferreira, and J.-M. G\'{e}rard,
Phys. Rev. B \textbf{75}, 073308 (2007).

\bibitem{Cook}R. J. Cook and H. J. Kimble,
Phys. Rev. Lett, \textbf{54}, 1023 (1985).












\end{thebibliography}
\end{document}